\documentclass[aps, pra, preprint,showpacs,superscriptaddress,amssymb,amsmath]{revtex4-1} 
\usepackage{latexsym}
\usepackage[pdftex]{graphicx}
\usepackage{bm}
\usepackage{epsfig}
\usepackage{dcolumn}

\begin{document}

\title{X-ray cross-correlation analysis of disordered systems: potentials and limitations}

\author{R.P. Kurta}
\affiliation{Deutsches Elektronen-Synchrotron DESY, Notkestra\ss e 85, D-22607 Hamburg, Germany}
\author{M. Altarelli}
\affiliation{European X-ray Free-Electron Laser Facility, Notkestra\ss e 85, D-22607, Hamburg, Germany}
\author{I.A. Vartanyants}
\email[Reference author: ]{ivan.vartaniants@desy.de}
\affiliation{Deutsches Elektronen-Synchrotron DESY, Notkestra\ss e 85, D-22607 Hamburg, Germany}
\affiliation{National Research Nuclear University, ``MEPhI'', 115409 Moscow, Russia}
\date{\today}

\begin{abstract}

Angular X-ray cross-correlation analysis (XCCA) is an approach to study the structure of disordered systems
using the results of x-ray scattering experiments. In this paper we summarize recent theoretical developments
related to the Fourier analysis of the cross-correlation functions. Results of our simulations demonstrate the application
of XCCA to two- and three-dimensional (2D and 3D) disordered systems of particles. We show that the structure of a single particle can be recovered using
x-ray data collected from a 2D disordered system of identical particles.
We also demonstrate that valuable structural information about the local structure of 3D systems,
inaccessible from a standard small-angle x-ray scattering experiment, can be resolved using XCCA.

\end{abstract}


\maketitle


\section{Introduction}

Correlation methods are widely used in the physics of disordered materials such as amorphous and glassy systems \cite{Stanley, Ngai}.
Recently it was suggested \cite{Wochner1, Altarelli, *[{Erratum: }]AltarelliErr, Kurta1, Kurta2,  Saldin1, Saldin2, Kirian1} that angular cross-correlations of scattered intensities in coherent diffraction experiment could provide essential structural information about the system that is not accessible by other scattering techniques.
The growing interest in the application of these methods is driven by two main factors. The first is its application to the physics of disordered or partially ordered systems, and the second is an attempt to solve the structure of biological molecules in their native environment.
Although ideas of applying cross-correlation techniques to study disordered systems go back to the work of Kam over 30 years ago \cite{Kam, Kam1}, new perspectives are opened by the emerging high power free-electron lasers (FELs) \cite{Ackermann, Emma, SCSS, Altarelli2}, which provide intense coherent femtosecond pulses of x-rays.

In study of disordered systems application of cross-correlation methods is driven by the hope to unveil the hidden symmetries and dynamics in a disordered collection of elements composing the system. This leads to the problem of understanding systems with  complicated correlation behavior between dynamical heterogeneity and medium-range order \cite{Elliott, Sheng, Treacy} in a large class of glass-forming liquids \cite{Tanaka3, Tanaka2, Tanaka1}.
One of the mysterious liquid systems whose structure is not fully understood is water. It exists in different forms known as low-density and high-density amorphous states \cite{Tulk}. The phase diagram of different forms is still under debate  \cite{Klug, Soper}.
In such systems the relevant task could be, for example, the detection of an $n$-fold symmetry axis of an individual molecular species in a liquid composed of such molecules. 
An identification of bond angles or local structure in an amorphous system, from the study of angular correlations of the diffracted intensity is another attractive topic of research
There are partly affirmative answers in the study of partially ordered quasi two-dimensional systems like liquid crystals \cite{Pindak} in which hexatic bond order can be detected from the angular intensity correlation.
 Angular correlations with pronounced periodic character that can be directly related to the the local symmetry of the system were also observed in colloidal suspensions \cite{Wochner1, Clark, Pusey}.

For biological systems it was proposed to use high power FEL pulses in so-called single particle coherent diffraction imaging experiments \cite{Neutze, Gaffney}.
In these experiments individual particles, for example macromolecules, are injected in the FEL beam and their diffraction patterns are measured in the far-field.
After the first experiments performed at LCLS in Stanford \cite{Hajdu}, it was realized that even high power FELs provide single pulse resolution only up to tens of nanometers for gigantic viruses. Extension of the technique to sub- nanometer resolution for biological systems with the size below 100 nm is still under development.
Clearly, it will be even more challenging to resolve molecules in their native liquid environment.
One of the possible ways to enhance the scattered signal is the illumination of a large number of particles by a single FEL pulse. An approach based on cross-correlation analysis could provide a tool to determine three-dimensional (3D) diffraction pattern to high resolution with the available FEL sources \cite{Saldin5, Saldin4, Starodub}.

\begin{figure*}[!htbp]
\centering
\includegraphics[width=1.0\textwidth]{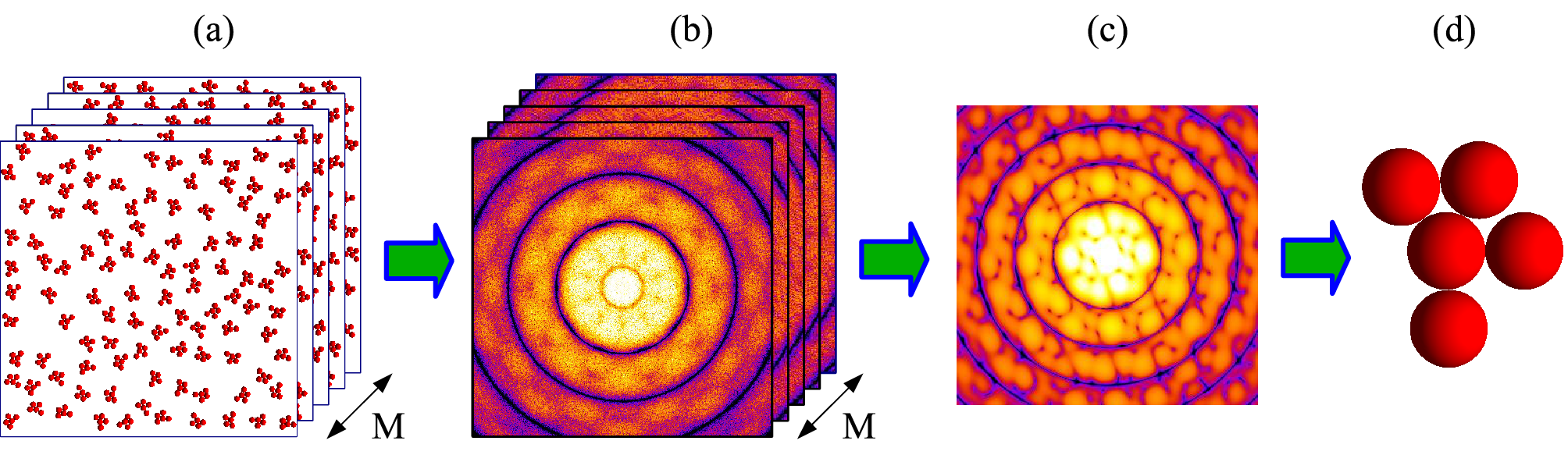}
\caption{
A concept of recovery of the structure of a single particle using x-ray scattering data from many particles. A large number $M$ of realizations of a disordered system (a) composed of many identical particles is used to collect $M$ diffraction patterns (b). X-ray cross-correlation analysis is applied to this x-ray data set to recover a diffraction pattern (c) corresponding to a single particle. The structure of a single particle (d) is determined applying phase retrieval algorithms to the recovered diffraction pattern (c).
\label{Fig:Intro}
}
\end{figure*}

Here, we discuss potentials of application of x-ray cross-correlation analysis (XCCA) to the study of a certain class of disordered systems.
Our model sample consists of $N$ identical non-interacting particles with orientational and positional disorder (see Fig.~\ref{Fig:GeomSampl}).
A coherent x-ray beam with a pulse duration shorter than the translational and rotational relaxation times of the system is scattered from the sample, and the diffraction pattern is measured in the far-field. The main concept of our approach based on application of XCCA to the study of disordered systems is depicted in Fig.~\ref{Fig:Intro}.
A large number of diffraction patterns is measured by successive pulses of the FEL beam with different realizations of the system.
We assume that the time between the pulses is sufficient that two sequential realizations of the system are completely uncorrelated.
XCCA is performed for each diffraction pattern in a sequence, and angular cross-correlation functions (CCFs) are averaged over the whole data set.
In this way structural information about an individual particle in the system can be obtained.

In this paper we summarize our recent theoretical results \cite{Altarelli, Kurta1, Kurta2, Kurta3} based on the analysis of the high-order cross-correlation functions.
It is organized as follows, in the next section the definition of the two-point and three-point CCFs, as well as their angular Fourier decomposition, is given. In the third section intensities scattered from our model system are defined.
In section four a detailed description of XCCA for 2D dilute and dense systems is provided, this is followed in the next section by the description of a more general case of scattering from 3D systems.
In section six major results of the recovery of the structure of an individual 2D cluster and analysis of the scattering from a 3D system of pentameric structures are discussed. This is followed by the conclusions and an overview section.

\section{Definition of the two- and three-point CCF{\scriptsize s} and their angular Fourier decomposition}

\begin{figure*}[!htbp]
\centering
\includegraphics[width=0.7\textwidth]{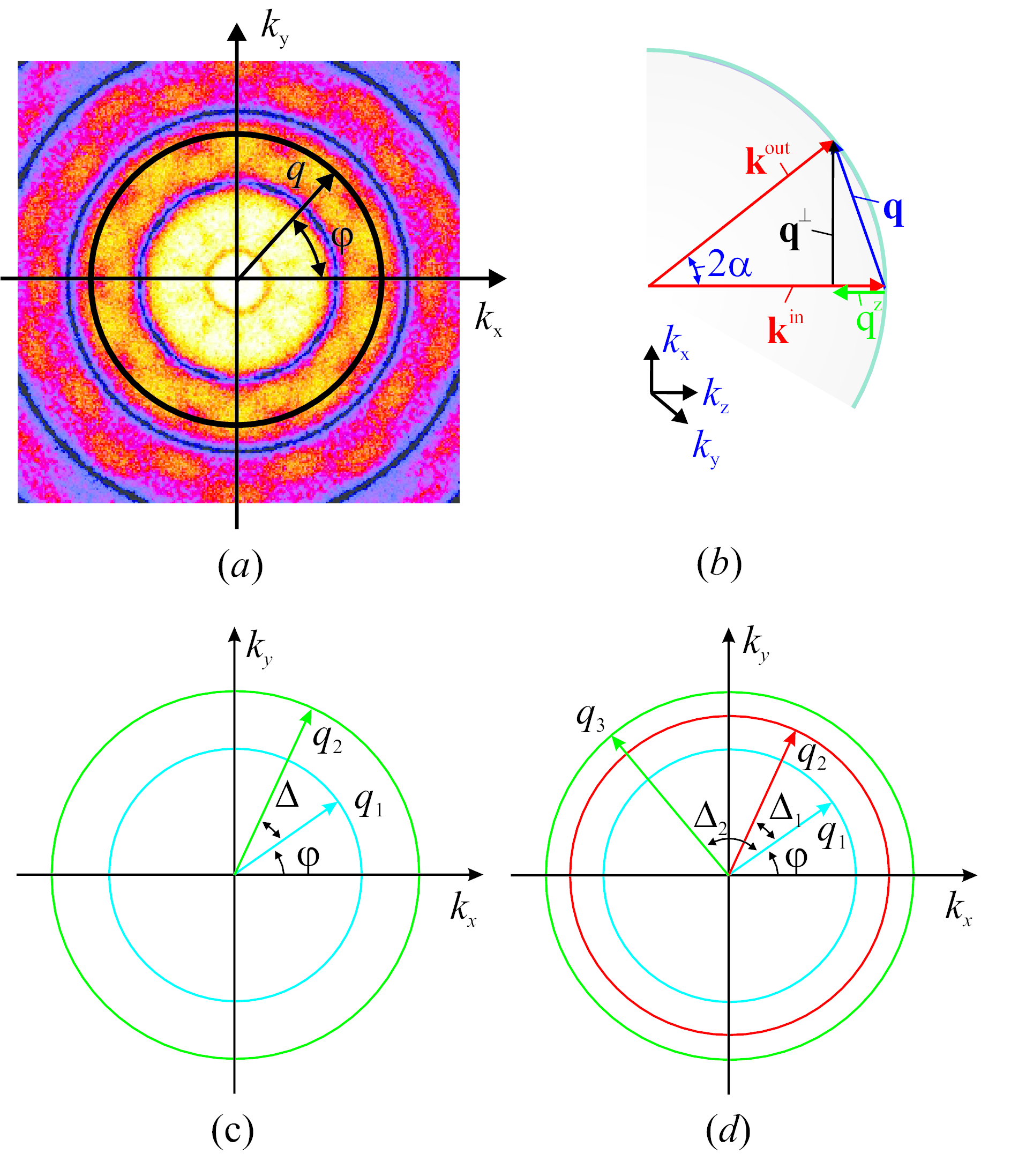}
\caption{
Scattering geometry in reciprocal space. (a)
Scattered intensity $I(\mathbf{ q})$ defined in the detector plane in the polar coordinate system, $\mathbf{q}=(q,\varphi)$.
(b) Ewald sphere construction. Here $\mathbf{ k}^{\text{in}}$ is the wavevector of the incident beam
directed along the $z$ axis, $\mathbf{ k}^{\text{out}}$ is the wavevector of the scattered wave with the
scattering angle $2\alpha$. The scattering vector $\mathbf{ q}=(\mathbf{ q}^{\perp},q^{z})$ is decomposed into two components that are perpendicular $\mathbf{ q}^{\perp}$ and parallel $q^{z}$ to the direction of the incident beam.
Definition of the momentum transfer vectors in the derivation of the two-point $C(q_{1},q_{2},\Delta)$ (c), and
three-point CCFs $C(q_{1},q_{2},q_{3},\Delta_{1},\Delta_{2})$ (d).
\label{Fig:Qvectors}
}
\end{figure*}

The two-point CCF defined for a single realization of a disordered system at two resolution rings $q_{1}$ and $q_{2}$ is given by \cite{Kam, Saldin, Altarelli} [see Fig.~\ref{Fig:Qvectors}(c)]
\begin{eqnarray}
C(q_{1},q_{2},\Delta)&&=\left\langle \widetilde{I}(q_{1},\varphi)\widetilde{I}(q_{2},\varphi+\Delta)\right\rangle_{\varphi},
\label{Cq1q2}
\end{eqnarray}
where $0\leq \Delta \leq2\pi$ is the angular coordinate, $\widetilde{I}(q,\varphi)=I(q,\varphi)- \left\langle I(q,\varphi)\right\rangle_{\varphi}$, is the intensity fluctuation function, $\langle f(\varphi) \rangle_{\varphi}=(1/2\pi)\int_{0}^{2\pi}f(\varphi)d\varphi$ denotes the average over the angle $\varphi$.

In a similar way, the three-point CCF for a single realization of a system is defined at three resolution rings $q_{1}$, $q_{2}$ and $q_{3}$ as \cite{Kurta2} [see Fig.~\ref{Fig:Qvectors}(d)],
\begin{equation}
 C(q_{1},q_{2},q_{3},\Delta_{1},\Delta_{2})=\left\langle \widetilde{I}(q_{1},\varphi)\widetilde{I}(q_{2},\varphi+\Delta_{1})\widetilde{I}(q_{3},\varphi+\Delta_{2})\right\rangle_{\varphi},
\label{Cq1q2q3}
\end{equation}
where $0\leq \Delta_{1} \leq2\pi$  and $0\leq \Delta_{2} \leq2\pi$ are the angular coordinates.

In practical applications one would need to consider
CCFs $\langle C(q_{1},q_{2},\Delta) \rangle_{M}$ and \linebreak
$\langle C(q_{1},q_{2},q_{3},\Delta_{1},\Delta_{2}) \rangle_{M}$ averaged over a sufficiently large number $M$ of diffraction patterns \cite{Kurta1},
\begin{subequations}
\begin{eqnarray}
&&\langle C(q_{1},q_{2},\Delta) \rangle_{M} = \frac{1}{M}\sum\limits_{m=1}^{M} \{C(q_{1},q_{2},\Delta)\}^{m},\label{AverCorr1}\\
&&\langle C(q_{1},q_{2},q_{3},\Delta_{1},\Delta_{2}) \rangle_{M} = \frac{1}{M}\sum\limits_{m=1}^{M} \{C(q_{1},q_{2},q_{3},\Delta_{1},\Delta_{2})\}^{m},\label{AverCorr2}
\end{eqnarray}
\end{subequations}
where $\{C(q_{1},q_{2},\Delta)\}^{m}$ and $\{C(q_{1},q_{2},q_{3},\Delta_{1},\Delta_{2})\}^{m}$ are the CCFs defined by Eqs.~(\ref{Cq1q2}) and (\ref{Cq1q2q3}) for the $m$-th realization of a disordered system.
The importance of this statistical averaging will be discussed in the following sections of the paper.

It is convenient to analyze the two-point CCF $C(q_{1},q_{2},\Delta) $ using a Fourier series decomposition
in the $(0,2\pi)$ interval \cite{Altarelli}
\begin{subequations}
\begin{eqnarray}
&&C(q_{1},q_{2},\Delta)=\sum\limits_{n=-\infty}^{\infty}C_{q_{1},q_{2}}^{n}e^{in\Delta},\label{Eq:Cq1q2_4}\\
&&C_{q_{1},q_{2}}^{n}=\frac{1}{2\pi}\int_{0}^{2\pi}C(q_{1},q_{2},\Delta)e^{-in\Delta}d\Delta.\label{Eq:Cq1q2n_1}
\end{eqnarray}
\end{subequations}
Here, $C_{q_{1},q_{2}}^{n}$ is the $n$-th component in the Fourier series expansion of $C(q_{1},q_{2},\Delta)$, and $C_{q_{1},q_{2}}^{0}=0$ at $n=0$ by definition [see Eq.~(\ref{Cq1q2})].
Substituting Eq.~(\ref{Cq1q2}) into Eq.~(\ref{Eq:Cq1q2n_1}) and applying the Fourier convolution theorem we get
\begin{eqnarray}
C_{q_{1},q_{2}}^{n}=I_{q_{1}}^{n\ast}\cdot I_{q_{2}}^{n}.
\label{Eq:Cq1q2n_2}
\end{eqnarray}
Here $I^{n}_{q}$ are the components of the Fourier expansion of the scattered intensity $I(q,\varphi)$ on the ring of radius $q$ [see Fig.~\ref{Fig:Qvectors}(a)],
\begin{subequations}
\begin{eqnarray}
&&I(q,\varphi)=\sum\limits_{n=-\infty}^{\infty} I^{n}_{q}e^{i n\varphi},\label{Eq:Iqn2}\\
&&I^{n}_{q}=\frac{1}{2\pi}\int_{0}^{2\pi}I(q,\varphi)e^{-in\varphi}d\varphi.\label{Eq:Iqn1}
\end{eqnarray}
\end{subequations}
Since scattered intensities are always real quantities, it is easy to show that $I^{-n}_{q}=I^{n \ast}_{q}$
and $C_{q_{1},q_{2}}^{-n}=C_{q_{1},q_{2}}^{n \ast}$.

In the specific case, when $q_{1}=q_{2}=q$, Eqs.~(\ref{Eq:Cq1q2_4}) and (\ref{Eq:Cq1q2n_2})
reduce to
\begin{subequations}
\begin{eqnarray}
&&C(q,\Delta)=2\sum\limits_{n=1}^{\infty} C_{q}^{n}\cos(n\Delta),
\label{Eq:Cq1q2_5q}\\
&&C_{q}^{n}=\mid I^{n}_{q}\mid^{2},\quad C_{q}^{n}\geq 0.
\label{Eq:Cq1q2n_2q}
\end{eqnarray}
\end{subequations}
According to (\ref{Eq:Cq1q2_5q}) a strong single cosine dependence of $C(q,\Delta)$ can be observed for those values of $q$,
at which one of the Fourier components $C_{q}^{n}$ significantly dominates over all others \cite{Wochner1}.
Such components can be related to the structure and symmetry of the system \cite{Altarelli, Kurta1}.

The Fourier series expansion of the three-point CCF  $C(q_{1},q_{2},q_{3},\Delta_{1},\Delta_{2})$ can be written as
\begin{subequations}
\begin{eqnarray}
&&C(q_{1},q_{2},q_{3},\Delta_{1},\Delta_{2})=
\sum\limits_{n_{1}=-\infty}^{\infty}
\sum\limits_{n_{2}=-\infty}^{\infty}
 C_{q_{1},q_{2},q_{3}}^{n_{1},n_{2}}e^{in_{1}\Delta_{1}}e^{in_{2}\Delta_{2}}, \label{Cq1q2q3FT1general1}\\
&&C_{q_{1},q_{2},q_{3}}^{n_{1},n_{2}}=\left(\frac{1}{2\pi}\right)^{2}\int_{0}^{2\pi}\int_{0}^{2\pi}
C(q_{1},q_{2},q_{3},\Delta_{1},\Delta_{2})e^{-in_{1}\Delta_{1}}e^{-in_{2}\Delta_{2}}d\Delta_{1}d\Delta_{2},\label{Cq1q2q3FT1general}
\end{eqnarray}
\end{subequations}
where $C_{q_{1},q_{2},q_{3}}^{n_{1},n_{2}}$ are the Fourier components of the three-point CCF, and $C_{q_{1},q_{2},q_{3}}^{n_{1},n_{2}}=0$ for $n_{1}= 0,\;n_{2}= 0$ and $n_{1} =-n_{2}$.
Substituting Eq.~(\ref{Cq1q2q3}) into Eq.~(\ref{Cq1q2q3FT1general}) one can get \cite{Kurta2}
\begin{eqnarray}
C_{q_{1},q_{2},q_{3}}^{n_{1},n_{2}}=I_{q_{1}}^{(n_{1}+n_{2})\ast}I_{q_{2}}^{n_{1}}I_{q_{3}}^{n_{2}},
\label{Eq:Cq1q2q3n1n2}
\end{eqnarray}
In general, Eq.~(\ref{Eq:Cq1q2q3n1n2}) determines a relation between three different Fourier components of intensity
 $I_{q}^{n}$ of the order $n_{1}$, $n_{2}$ and $n_{1}+n_{2}$, defined on three resolution rings, $q_{1},q_{2}$ and $q_{3}$.

In practical applications one would need to consider the Fourier spectra
$\langle C_{q_{1},q_{2}}^{n} \rangle_{M}$ and  $\langle C_{q_{1},q_{2},q_{3}}^{n_{1},n_{2}} \rangle_{M}$ averaged over a large number $M$ of diffraction patterns \cite{Kurta1, Kurta2},
\begin{subequations}
\begin{eqnarray}
&&\langle C_{q_{1},q_{2}}^{n} \rangle_{M} = 1/M \sum\limits_{m=1}^{M} \{C_{q_{1},q_{2}}^{n}\}^{m},\label{AverCorrSpec1}\\
&&\langle C_{q_{1},q_{2},q_{3}}^{n_{1},n_{2}} \rangle_{M} = 1/M \sum\limits_{m=1}^{M} \{C_{q_{1},q_{2},q_{3}}^{n_{1},n_{2}}\}^{m},\label{AverCorrSpec2}
\end{eqnarray}
\end{subequations}
where $\{C_{q_{1},q_{2}}^{n}\}^{m}$ and $\{C_{q_{1},q_{2},q_{3}}^{n_{1},n_{2}}\}^{m}$ are the Fourier components of the CCFs $\{C(q_{1},q_{2},\Delta)\}^{m}$
and $\{C(q_{1},q_{2},q_{3},\Delta_{1},\Delta_{2})\}^{m}$ respectively, defined for the $m$-th realization of the system [see Eqs.~(\ref{AverCorr1}) and (\ref{AverCorr2})].

\section{Scattering from a disordered system of identical particles}

\begin{figure*}[!htbp]
\centering
\includegraphics[width=0.8\textwidth]{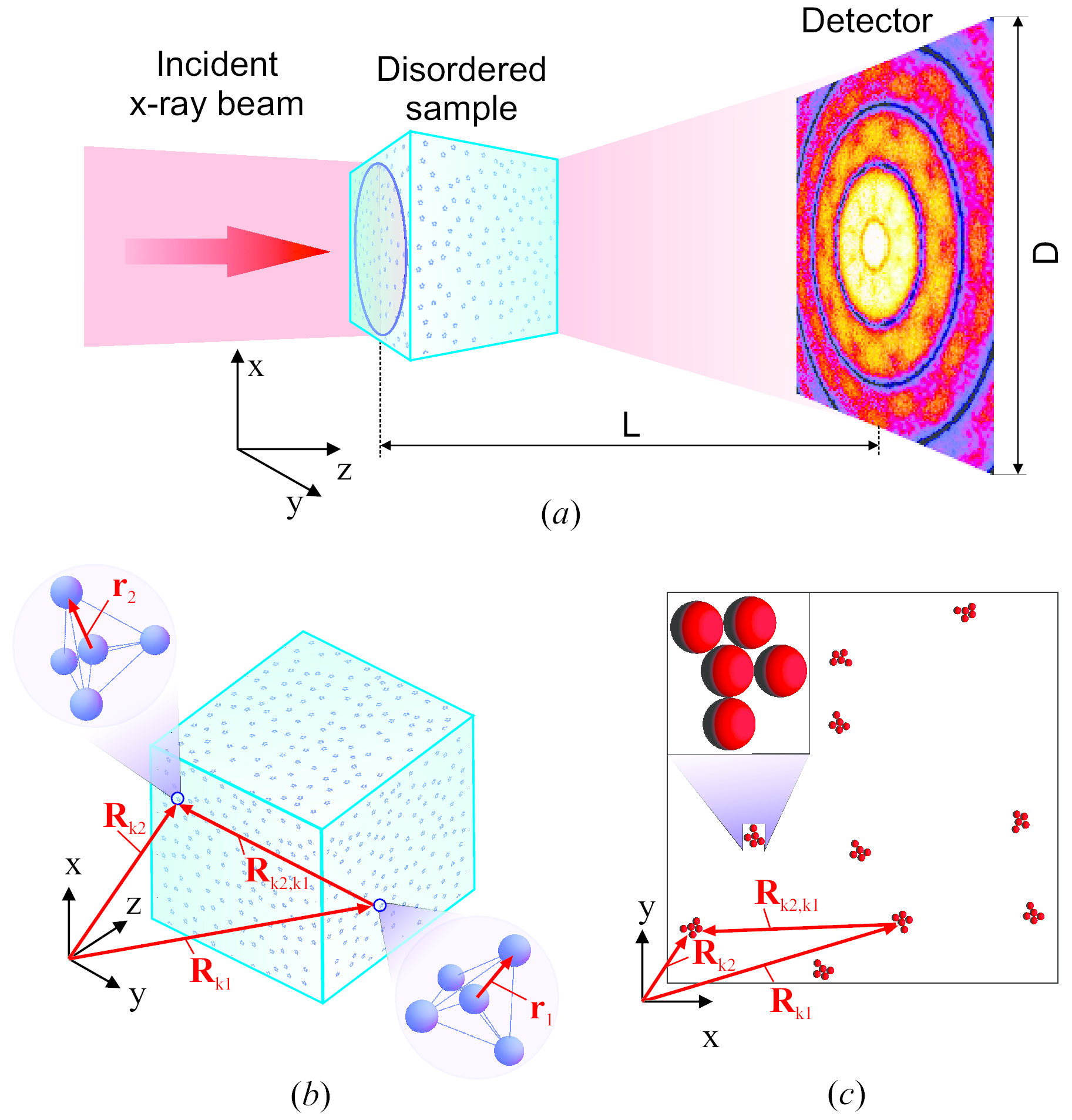}
\caption{
(a) Geometry of the diffraction experiment. A coherent x-ray beam illuminates a disordered sample and produces a diffraction pattern on a detector. The direction of the incident beam is defined along the z axis of the coordinate system. (b)  A disordered 3D sample composed of tetrahedral pentamers. All clusters have random position and orientation in the 3D space. (c) A disordered 2D sample composed of asymmetric clusters. All clusters have random position and orientation in the 2D plane.
\label{Fig:GeomSampl}
}
\end{figure*}

We consider a scattering experiment in transmission geometry as shown
in Fig.~\ref{Fig:GeomSampl}(a).
A coherent x-ray beam scatters from a disordered sample,
and a speckle pattern is measured on the detector in the far-field regime. In our simulations we consider
a kinematical scattering approximation.
As a general model system we assume a 3D
sample consisting of $N$ identical particles with random positions and orientations.
This model includes a variety of systems, for example, clusters or molecules in the gas phase, local structures formed
in colloidal systems, viruses or complex biological systems in solution, etc.

The amplitude $A_{k}(\mathbf{ q})$ scattered from the $k$-th particle at the momentum transfer vector $\mathbf{q}$ can be defined as \cite{Altarelli},
\begin{equation}
A_{k}(\mathbf{q})=\int\rho_{k}(\mathbf{r})e^{i\mathbf{ q\cdot r}}d\mathbf{ r},\label{Eq:Aq}
\end{equation}
where $\rho_{k}(\mathbf{r})$ is an electron density of the $k$-th particle at
the position $\mathbf{R}_{k}$ [see Fig.~\ref{Fig:GeomSampl}(b)] and the integration is performed over the volume of the particle.
Using Eq.~(\ref{Eq:Aq}) the intensity $I(\mathbf{ q})$ coherently scattered from a disordered sample consisting of $N$ particles
is given by
\begin{eqnarray}
I(\mathbf{ q}) & = & \sum\limits _{k_{1},k_{2}=1}^{N}e^{i\mathbf{ q}\cdot\mathbf{ R}_{k_{2},k_{1}}}A_{k_{1}}^{\ast}(\mathbf{ q})A_{k_{2}}(\mathbf{ q})\nonumber \\
& = & \sum\limits _{k_{1},k_{2}=1}^{N}
\int\int\rho_{k_{1}}^{\ast}(\mathbf{ r}_{1})\rho_{k_{2}}(\mathbf{ r}_{2})e^{i\mathbf{ q}\cdot\mathbf{ R}_{k_{2},k_{1}}^{21}}d\mathbf{ r}_{1}d\mathbf{ r}_{2},\label{Intens1}
\end{eqnarray}
where the double summation is performed over all $N$ particles, and the integration is performed over the volume of the $k_{i}$-th particle $(i=1,2)$.
Here, the following notation for the radius vectors is used, $\mathbf{ R}_{k_{2},k_{1}}^{21}=\mathbf{ R}_{k_{2},k_{1}}+\mathbf{ r}_{21}$,
where $\mathbf{ R}_{k_{2},k_{1}}=\mathbf{ R}_{k_{2}}-\mathbf{ R}_{k_{1}}$
is the vector connecting two different particles $k_{1}$ and $k_{2}$, and $\mathbf{ r}_{21}=\mathbf{ r}_{2}-\mathbf{ r}_{1}$, where
the vectors $\mathbf{ r}_{1}$ and $\mathbf{ r}_{2}$ define the positions of scatterers (for example, colloidal spheres or atoms) inside the particles $k_{1}$ and $k_{2}$ respectively [see Fig.~\ref{Fig:GeomSampl}(b),(c)].

In the case of a partially coherent illumination and a dilute disordered system when the mean distance between the particles is larger than the coherence length of the incoming beam,
the inter-particle correlations due to coherent interference of scattered amplitudes from the individual particles in
Eq.~(\ref{Intens1}) can be neglected. In these conditions, the total scattered intensity $I(\mathbf{q})$ can be represented as a sum of intensities $I_{k}(\mathbf{q})=\left\vert A_{k}(\mathbf{ q}) \right\vert^{2}$ corresponding to individual particles in the system
\begin{equation}
I(\mathbf{q})=\sum\limits_{k=1}^{N} I_{k}(\mathbf{q}).
\label{Iincoh1}
\end{equation}

\section{2D disordered systems, small angle scattering}

In this section we consider a particular case of a 2D system in a small angle scattering geometry when all vectors are defined in a 2D plane [see Fig.~\ref{Fig:GeomSampl}(c)].
It was shown \cite{Altarelli}, that in this case only even ($n=2l, l=1,2,3,\dots$) Fourier components of the intensity $I^{n}_{q}$ can have non-zero values.

\subsection{Dilute systems}

The intensity $I_{\psi_{0}}(\mathbf{q})$ scattered from a single particle in some reference orientation
$\psi_{0}$ \footnote{Without loss of generality we fix the reference orientation to $\psi_{0}=0$.} 
is related to the electron density of the particle $\rho_{\psi_{0}}(\mathbf{r})$
through its scattered amplitude [Eq.~(\ref{Eq:Aq})] as $I_{\psi_{0}}(\mathbf{q})=\left|A_{\psi_{0}}(\mathbf{q})\right|^{2}$.
%
%
Similar to $I(\mathbf{q})$ [see Eqs.~(\ref{Eq:Iqn2}) and (\ref{Eq:Iqn1})], the intensity $I_{\psi_{0}}(\mathbf{q})\equiv I_{\psi_{0}}(q,\varphi)$
can be represented as an angular Fourier series expansion,
\begin{eqnarray}
I_{\psi_{0}}(q,\varphi)=\sum\limits_{n=-\infty}^{\infty} I_{q,\psi_{0}}^{n}e^{i n\varphi},\label{IFSeries1}
\end{eqnarray}
where $I_{q,\psi_{0}}^{n}$ are the Fourier components of $I_{\psi_{0}}(q,\varphi)$.

For a dilute 2D system of identical particles, the intensity $I_{\psi_{k}}(q,\varphi)$ scattered from a particle in an arbitrary orientation $\psi_{k}$
is related to the intensity $I_{\psi_{0}}(q,\varphi)$ scattered from a particle in the reference orientation $\psi_{0}$
as $I_{\psi_{k}}(q,\varphi)=I_{\psi_{0}}(q,\varphi-\psi_{k})$.
Applying the shift theorem for the Fourier transforms \cite{Oppenheim} we obtain for the corresponding Fourier components of the intensities, $I_{q,\psi_{k}}^{n}=I_{q,\psi_{0}}^{n}\exp(-i n\psi_{k})$.
Using these relations we can write for the Fourier components $I_{q}^{n}$ of the intensity $I(q,\varphi)$ scattered from $N$ particles [Eq.~(\ref{Iincoh1})]
\begin{equation}
 I_{q}^{n}=I_{q,\psi_{0}}^{n}\sum_{k=1}^{N}e^{-i n\psi_{k}}=I_{q,\psi_{0}}^{n}\mathbf{A}_{n},
\label{FCrelIncoh}
\end{equation}
where $\mathbf{A}_{n}=\sum_{k=1}^{N}\exp(-i n\psi_{k})$ is a random phasor sum \cite{Goodman2}.

The Fourier component $I_{q,\psi_{0}}^{n}$ is related to the projected electron density $\widetilde{\rho}_{\psi_{0}}(\mathbf{r})$ of the particle as \cite{Altarelli,Kurta1}
\begin{equation}
I_{q,\psi_{0}}^{n}=\int\int \widetilde{\rho}_{\psi_{0}}^{\ast}(\mathbf{ r}_{1} ) \widetilde{\rho}_{\psi_{0}}(\mathbf{ r}_{2} )J_{n}(q
|\mathbf{ r}_{21} |)e^{-in\phi_{\mathbf{ r}_{21} }} d\mathbf{ r}_{1} d\mathbf{ r}_{2},
\label{LocStr1}
\end{equation}
where $\phi_{\mathbf{ r}_{21} }$ is the angle of the vector $\mathbf{ r}_{21}$ in the detector plane,
 $J_{n}(\rho)$ is the Bessel function of the first kind of integer order $n$, and the integration is performed over the area of a particle.
According to the structure of $I_{q,\psi_{0}}^{n}$ its value strongly depends on the symmetry of a particle and
determines selection rules \cite{Altarelli,Kurta1} for the values $n$ of non-zero Fourier components $C_{q_{1},q_{2}}^{n}$.
These selection rules can be used for the identification of the symmetry of particles in dilute systems.
For example, for a cluster with 5-fold symmetry only $n=10l,\;(l=1,2...)$ will give a non-zero contribution to the Fourier components of CCFs.

Substituting Eq.~(\ref{FCrelIncoh}) in Eq.~(\ref{Eq:Cq1q2n_2}), in the limit of dilute systems we have for the Fourier components $C_{q_{1},q_{2}}^{n}$ of the CCF the following expression
\begin{eqnarray}
C_{q_{1},q_{2}}^{n}=I_{q_{1},\psi_{0}}^{n\ast}I_{q_{2},\psi_{0}}^{n} \left\vert {\mathbf A}_{n}\right\vert^2,
\label{Cq1q2nDilute}
\end{eqnarray}

The statistical behavior of ${\mathbf A}_{n}$ has been analyzed for different angular distributions of orientations of particles in the system [see Refs. \cite{Altarelli, Kurta1}]. It is clear, that
in the case of a completely oriented system of particles (all $\psi_{k}=0$) the square amplitude of the random phasor sum $\left\vert {\mathbf A}_{n}\right\vert^2$ is equal to $N$ and  $C_{q_{1},q_{2}}^{n}=NI_{q_{1},\psi_{0}}^{n\ast}I_{q_{2},\psi_{0}}^{n}$.
In the case of a uniform distribution of orientations of particles $\left\vert {\mathbf A}_{n}\right\vert^2$
fluctuates around its mean value $\langle \left\vert {\mathbf A}_{n}\right\vert^2 \rangle=N$ with the standard deviation $\sigma_{\left\vert {\mathbf A}_{n}\right\vert^2}=N$.
Averaging the Fourier components $C_{q_{1},q_{2}}^{n}$ over a large number $M$ of diffraction patterns decreases these fluctuations and leads to the following asymptotic result \cite{Altarelli, Kurta1}
\begin{equation}
\left\langle C_{q_{1},q_{2}}^{n}\right\rangle_{M} = I_{q_{1},\psi_{0}}^{n\ast}I_{q_{2},\psi_{0}}^{n} \cdot \left\langle | \mathbf{A}_{n} |^{2}\right\rangle_{M} \underset{M\rightarrow\infty}{\longrightarrow}  NI_{q_{1},\psi_{0}}^{n\ast}I_{q_{2},\psi_{0}}^{n}.
\label{Cq1q2Iq1Iq2A}
\end{equation}
%
%
where $\left\langle \dots \right\rangle_{M}$ denotes statistical averaging over $M$  diffraction patterns.
Importantly, the ensemble-averaged Fourier components $\left\langle C_{q_{1},q_{2}}^{n}\right\rangle_{M}$ converge to a scaled product of the two Fourier components of intensity $I_{q_{1},\psi_{0}}^{n\ast}$  and $I_{q_{2},\psi_{0}}^{n}$ associated with a single particle.

The amplitudes $|I_{q,\psi_{0}}^{n}|$ and phases $\phi_{q,\psi_{0}}^{n}$ (for $n\neq0$) of the Fourier components $I_{q,\psi_{0}}^{n}=|I_{q,\psi_{0}}^{n}|\exp(i \phi_{q,\psi_{0}}^{n})$ associated with a single particle can be determined using Eq.~(\ref{Cq1q2Iq1Iq2A}) \cite{Kurta2}. This equation determines the phase difference between two Fourier components
$I_{q_{1},\psi_{0}}^{n}$ and $I_{q_{2},\psi_{0}}^{n}$ of the same order $n$, defined at two different resolution rings $q_{1}$ and $q_{2}$,
\begin{equation}
\arg[\left\langle C_{q_{1},q_{2}}^{n}\right\rangle_{M}]=\phi_{q_{2},\psi_{0}}^{n}-\phi_{q_{1},\psi_{0}}^{n}.
\label{PhRel1}
\end{equation}

Similar to the Fourier components of the two-point CCF, the Fourier components of the three-point CCF Eq.~(\ref{Eq:Cq1q2q3n1n2})
can be expressed in the limit of a dilute system as \cite{Kurta2}
\begin{equation}
 C_{q_{1},q_{2},q_{3}}^{n_{1},n_{2}}=I_{q_{1},\psi_{0}}^{(n_{1}+n_{2})\ast} I_{q_{2},\psi_{0}}^{n_{1}} I_{q_{3},\psi_{0}}^{n_{2}}
\cdot \mathbf{A}_{n_{1},n_{2}} ,
\label{Cq1q2q3FT2}
\end{equation}
where $\mathbf{A}_{n_{1},n_{2}}=\sum_{i,j,k=1}^{N}\exp\{i[(n_{1}+n_{2})\psi_{i}-n_{1}\psi_{j}-n_{2}\psi_{k}]\}$ is another random phasor sum. 
Our analysis shows \cite{Kurta2} that in the case of a uniform distribution of orientations of $N$ particles the statistical average $\left\langle \mathbf{A}_{n_{1},n_{2}} \right\rangle_{M}$ converges to $N$ for a sufficiently large number $M$ of diffraction patterns
\begin{equation}
\left\langle C_{q_{1},q_{2},q_{3}}^{n_{1},n_{2}}\right\rangle_{M} \underset{M\rightarrow\infty}{\longrightarrow} N I_{q_{1},\psi_{0}}^{(n_{1}+n_{2})\ast} I_{q_{2},\psi_{0}}^{n_{1}} I_{q_{3},\psi_{0}}^{n_{2}}.
\label{Cq1q2q3FT3}
\end{equation}
An important result of Eq.~(\ref{Cq1q2q3FT3}) is that the ensemble-averaged Fourier components $\left\langle C_{q_{1},q_{2},q_{3}}^{n_{1},n_{2}}\right\rangle_{M}$ converge to a scaled product of three Fourier components of intensity $I_{q_{1},\psi_{0}}^{(n_{1}+n_{2})\ast}$, $I_{q_{2},\psi_{0}}^{n_{1}}$,  and $I_{q_{3},\psi_{0}}^{n_{2}}$ associated with a single particle. Eq.~(\ref{Cq1q2q3FT3}) also provides the following phase relation,
\begin{equation}
\arg[\left\langle C_{q_{1},q_{2},q_{3}}^{n_{1},n_{2}} \right\rangle_{M}]=\phi_{q_{2},\psi_{0}}^{n_{1}}+\phi_{q_{3},\psi_{0}}^{n_{2}}-\phi_{q_{1},\psi_{0}}^{(n_{1}+n_{2})}.
\label{PhRel2}
\end{equation}
This equation determines the phase difference between three Fourier components $I_{q_{1},\psi_{0}}^{(n_{1}+n_{2})}$, $I_{q_{2},\psi_{0}}^{n_{1}}$, and $I_{q_{3},\psi_{0}}^{n_{2}}$ of different order $n$ defined on three resolution rings. If $n_{1}=n_{2}=n$ and $n_{3}=2n$, equation (\ref{PhRel2}) reduces to a particular form, giving the phase relation between Fourier components of only two different orders $n$ and $2n$.
Phase relations (\ref{PhRel1}) and (\ref{PhRel2}) can be used to determine the phases of the complex Fourier components $I_{q,\psi_{0}}^{n}$
using measured CCFs from a disordered system of particles \cite{Kurta2}.

\subsection{Dense systems}

In the case of a dense system, when the average distance between particles is of the order of the size of a single cluster,
the Fourier components $I^{n}_{q}$ of the intensity $I(\mathbf{ q})$ [Eq.~(\ref{Intens1})] can contain a substantial inter-particle contribution.
In this case $I^{n}_{q}$ can be presented as a sum of two terms,
\begin{eqnarray}
I^{n}_{q} =I^{n}_{\rm{part}}(q)+I^{n}_{\rm{int-part}}(q),
\label{Inq}
\end{eqnarray}
where $I^{n}_{\rm{part}}(q)$ is attributed to a single particle structure discussed above, and $I^{n}_{\rm{int-part}}(q)$ is defined by the inter-particle correlations \cite{Altarelli,Kurta1}.
In a 2D system these two terms are \cite{Altarelli,Kurta1}
\begin{subequations}
\begin{eqnarray}
&&I^{n}_{\rm{part}}(q)=I_{q,\psi_{0}}^{n}\mathbf{A}_{n},\label{Lnk}\\
&&I^{n}_{\rm{int-part}}(q) =2\sum\limits_{k_{2}>k_{1}} \int\int \widetilde{\rho}_{k_{1}}^{\ast}(\mathbf{ r}_{1})\widetilde{\rho}_{k_{2}}(\mathbf{ r}_{2})
J_{n}(q|\mathbf{ R}_{k_{2},k_{1}}^{21}|)e^{-in\phi_{\mathbf{ R}_{k_{2},k_{1}}^{21}}}d\mathbf{ r}_{1} d\mathbf{ r}_{2},
\label{Ln_k1k2}
\end{eqnarray}
\end{subequations}
where $\phi_{\mathbf{ R}_{k_{2},k_{1}}^{21}}$ is the angle of the vector $\mathbf{ R}_{k_{2},k_{1}}^{21}$ in the sample plane.

Taking into account both terms of Eq.~(\ref{Inq}), the Fourier components of the two-point CCF
for a single realization of a system (\ref{Eq:Cq1q2n_2}) can be written as the following sum of four terms
\begin{equation}
C_{q_{1},q_{2}}^{n}=I_{q_{1}}^{n\ast}\cdot I_{q_{2}}^{n}=S^{n}_{1}+(S^{n}_{2}+S^{n}_{3})+S^{n}_{4},
\label{Cq1q2n_2}
\end{equation}
where
\begin{subequations}
\begin{eqnarray}
S^{n}_{1}&&= I^{n\ast}_{\rm{part}}(q_{1})I^{n}_{\rm{part}}(q_{2}),\label{Cq1q2n_T1}\\
S^{n}_{2}+S^{n}_{3}&&= I^{n\ast}_{\rm{part}}(q_{1})I^{n}_{\rm{int-part}}(q_{2})+I^{n\ast}_{\rm{int-part}}(q_{1})I^{n}_{\rm{part}}(q_{2}) ,\label{Cq1q2n_T23}\\
S^{n}_{4}&&=I^{n\ast}_{\rm{int-part}}(q_{1})I^{n}_{\rm{int-part}}(q_{2}).\label{Cq1q2n_T4}
\end{eqnarray}
\end{subequations}
In Eq.~(\ref{Cq1q2n_2}) the term $S^{n}_{1}$ is defined only by the structure of a particle, whereas the terms $S^{n}_{2}$, $S^{n}_{3}$ and
$S^{n}_{4}$ contain contributions from the inter-particle correlations due to the second term in Eq.~(\ref{Inq}).
A similar expansion can be performed for the Fourier components of the three-point CCF. The influence of inter-particle correlations on CCF's measured on the same resolution ring were discussed in detail in Ref. \cite{Kurta1}.

\section{3D systems, wide angle scattering}

In our previous discussion of x-ray scattering on 2D systems, we have seen that only even Fourier components of the CCFs have non-zero values.
In the case of 3D systems non-zero odd Fourier components can be also present when scattering to high angles is considered, due to Ewald sphere curvature effects.

In general, the scattering vector $\mathbf{ q}=(\mathbf{ q}^{\perp},q^{z})$ can be decomposed
into two components: $\mathbf{ q}^{\perp}$ that is perpendicular,
and $q^{z}$ that is parallel to the direction of the incident beam [see Fig.~\ref{Fig:Qvectors}(b)].
We also define the perpendicular
$\mathbf{ R}_{k_{2},k_{1}}^{\perp 21}=\mathbf{ R}_{k_{2},k_{1}}^{\perp}+\mathbf{ r}_{21}^{\perp}$,\
and the $z$-components
$Z_{k_{2},k_{1}}^{21}=Z_{k_{2},k_{1}}+z_{21}$
of the radius vectors (see Figs.~\ref{Fig:GeomSampl}(a) and \ref{Fig:Qvectors}).
Using these notations we can write Eq.~(\ref{Intens1}) in the following form
\begin{equation}
I(\mathbf{ q})= \sum\limits _{k_{1},k_{2}=1}^{N}e^{-iq^{z}\cdot Z_{k_{2},k_{1}}}
 \int\int\widetilde{\rho}_{k_{1}}^{\ast}(\mathbf{ r}_{1}^{\perp},q^{z})\widetilde{\rho}_{k_{2}}(\mathbf{ r}_{2}^{\perp},q^{z})e^{i\mathbf{ q}^{\perp}\cdot \mathbf{ R}_{k_{2},k_{1}}^{\perp 21}}d\mathbf{ r}^{\perp}_{1}d\mathbf{ r}^{\perp}_{2}.
 \label{Intens2}
 \end{equation}
Here we introduced a modified complex valued electron density function, defined as
\begin{equation}
\widetilde{\rho}_{k_{i}}(\mathbf{ r}_{i}^{\perp},q^{z})=\int\rho_{k_{i}}(\mathbf{ r}_{i}^{\perp},z)e^{-iq^{z}z}dz.
\label{Eq:Ro_2}
\end{equation}

In the case of wide angle scattering, the effect of the Ewald sphere curvature [see Fig.~\ref{Fig:Qvectors}(b)], which manifests itself
by the presence of the exponential factors $e^{-iq^{z}\cdot Z_{k_{2},k_{1}}}$ and $e^{-iq^{z}z}$ in Eqs.~(\ref{Intens2}) and (\ref{Eq:Ro_2}), may become important.
This effect can break the scattering symmetry of a diffraction pattern, characteristic for the scattering on a positive valued electron density (Friedel's law) and
may reveal additional symmetries that can be hidden in the small angle scattering case. A wide angle scattering geometry may become important for scattering on atomic systems
with local interatomic distances of the order of few \AA ngstroms.

For simplicity we will consider here a 3D system consisting of
particles composed of identical scatterers. The modified electron density of a particle according to Eq.~(\ref{Eq:Ro_2}) can be defined in the following form
\begin{equation}
\widetilde{\rho}_{k}(\mathbf{ r}^{\perp},q^{z})=f(q)
\sum\limits _{i=1}^{N_{s}}\delta(\mathbf{ r}^{\perp}-\mathbf{ r}^{\perp}_{i})e^{-iq^{z}z_{i}},
\label{Eq:Ro_4}
\end{equation}
where $f(q)$ is a form-factor of a scatterer, and $N_{s}$ is a number of scatterers in the cluster. The coordinates $(\mathbf{ r}^{\perp}_{i},z_{i})$ define the position of the $i$-th scatterer inside the $k$-th cluster.
Using this definition and performing a Fourier transformation of Eq.~(\ref{Intens2}) we obtain \cite{Altarelli}
 \begin{eqnarray}
{ I}^{n}(q^{\perp} ,q^{z}) =(i)^{n}\left|f(q)\right|^{2}
\sum\limits _{k_{1},k_{2}=1}^{N}
\sum\limits _{l,m=1}^{N_{s}}
 e^{-iq^{z}Z_{k_{2},k_{1}}^{ml}}
J_{n}(\vert \mathbf{ q}^{\perp} \vert\cdot|\mathbf{ R}_{k_{2},k_{1}}^{\perp ml}|)e^{-in\phi_{\mathbf{ R}_{k_{2},k_{1}}^{\perp ml}}},
\label{Eq:In_k1k2_3}
\end{eqnarray}
where the summation over index $l$ is performed over the positions of scatterers in the cluster $k_{1}$, and
the summation over index  $m$ is performed over the positions of scatterers in the cluster $k_{2}$.
We note here that due to the property of the Bessel functions [$J_{n}(0)=0$ for $n\neq 0$] the terms with  $k_{1}=k_{2}$ and $l=m$ are equal to zero.
Taking into account
that the terms with interchanged indices, i.e. $k_{1}, k_{2}$ and $k_{2}, k_{1}$, as well as $l, m$ and $m, l$,
differ from each other by a change of the sign of $Z_{k_{2},k_{1}}^{ml}$ and by an additional factor $(-1)^{n}$,
which arises due to the change of the phase $\phi_{\mathbf{ R}_{k_{2},k_{1}}^{\perp ml}}=\phi_{\mathbf{ R}_{k_{1},k_{2}}^{\perp lm}}+\pi$,
we have for $\textit{even}$ values of $n$ in Eq.~(\ref{Eq:In_k1k2_3}) \cite{AltarelliErr}
\begin{eqnarray}
{I}^{n}(q^{\perp} ,q^{z}) =2(i)^{n} \left|f(q)\right|^{2}
&&\lbrack\sum\limits_{\begin{subarray}{c}
1\leq k_{1} \leq N\\
k_{1}< k_{2} \leq N
\end{subarray}}
\sum\limits_{\begin{subarray}{c}
1\leq l \leq N_{s}\\
1\leq m \leq N_{s}
\end{subarray}}
\cos{(q^{z}Z_{k_{2},k_{1}}^{ml})}
J_{n}(\vert \mathbf{ q}^{\perp} \vert\cdot|\mathbf{ R}_{k_{2},k_{1}}^{\perp ml}|)e^{-in\phi_{\mathbf{ R}_{k_{2},k_{1}}^{\perp ml}}} \nonumber\\
+&&\sum\limits_{
1\leq k \leq N}
\sum\limits_{\begin{subarray}{c}
1\leq l \leq N_{s}\\
l\leq m \leq N_{s}
\end{subarray}}
\cos{(q^{z}Z_{k}^{ml})}
J_{n}(\vert \mathbf{ q}^{\perp} \vert\cdot|\mathbf{ R}_{k}^{\perp ml}|)e^{-in\phi_{\mathbf{ R}_{k}^{\perp ml}}}\rbrack
\label{Eq:Ln_k1k2_4}
\end{eqnarray}
and for $\textit{odd}$ values of $n$:

\begin{eqnarray}
{I}^{n}(q^{\perp} ,q^{z}) =-2(i)^{n+1} \left|f(q)\right|^{2}
&&[\sum\limits_{\begin{subarray}{c}
1\leq k_{1} \leq N\\
k_{1}< k_{2} \leq N
\end{subarray}}
\sum\limits_{\begin{subarray}{c}
1\leq l \leq N_{s}\\
1\leq m \leq N_{s}
\end{subarray}}
\sin{(q^{z}Z_{k_{2},k_{1}}^{ml})}
J_{n}(\vert \mathbf{ q}^{\perp} \vert\cdot|\mathbf{ R}_{k_{2},k_{1}}^{\perp ml}|)e^{-in\phi_{\mathbf{ R}_{k_{2},k_{1}}^{\perp ml}}} \nonumber\\
+&&\sum\limits_{
1\leq k \leq N}
\sum\limits_{\begin{subarray}{c}
1\leq l \leq N_{s}\\
l\leq m \leq N_{s}
\end{subarray}}
\sin{(q^{z}Z_{k}^{ml})}
J_{n}(\vert \mathbf{ q}^{\perp} \vert\cdot|\mathbf{ R}_{k}^{\perp ml}|)e^{-in\phi_{\mathbf{ R}_{k}^{\perp ml}}}],
\label{Eq:Ln_k1k2_3}
\end{eqnarray}
where $\mathbf{ R}_{k}^{\perp ml}=\mathbf{ R}_{k,k}^{\perp ml}$ and $Z_{k}^{ml}= Z_{k,k}^{ml}$.
From the performed analysis we can see that, due to the curvature of the Ewald sphere (non-zero $q^{z}$ component),
we obtain non-zero odd Fourier components of the CCF when scattering from a 3D system. These components become negligibly small for experimental conditions corresponding to a flat Ewald sphere, considered in the previous section.

In the case of a dilute sample, the Fourier components of intensity defined in Eq.~(\ref{Eq:Ro_4}) reduce to
 \begin{eqnarray}
{ I}^{n}(q^{\perp} ,q^{z}) =(i)^{n}\left|f(q)\right|^{2}
\sum\limits _{k=1}^{N}
\sum\limits _{l,m=1}^{N_{s}}
 e^{-iq^{z}z_{k}^{ml}}
J_{n}(\vert \mathbf{ q}^{\perp} \vert\cdot|\mathbf{ r}_{k}^{\perp ml}|)e^{-in\phi_{\mathbf{ r}_{k}^{\perp ml}}}.
\label{Eq:In_k1k2_3dil}
\end{eqnarray}

\section{Results and discussion}

In this section we provide two examples of application of XCCA for investigation of the structural properties of disordered systems.

\subsection{Recovery of the structure of an individual 2D particle}

In the pioneering work of Kam \cite{Kam, Kam1} it was proposed to determine the structure of a single particle using scattered intensity from many identical particles in solution.
However, this approach, based on a spherical harmonics expansion of the scattered amplitudes, was not fully explored until now.
Recently, it has been revised theoretically \cite{Saldin, Saldin1, Saldin4} and experimentally \cite{Saldin2}. The possibility to recover the structure of individual particle was demonstrated in systems in two dimensions  \cite{Saldin,Saldin1,Saldin2} and three dimensions  \cite{Saldin4} using additional {\it a priori} knowledge on the symmetry of the particles. Unfortunately, these approaches, based on optimization routines \cite{Saldin,Saldin4} and iterative techniques \cite{Saldin1, Saldin2}, do not guarantee the uniqueness of the recovered structure unless strong constraints are applied. Here we provide a brief overview of our recently developed approach \cite{Kurta2} that enables a direct reconstruction of the single-particle intensity distribution using an algebraic formalism of two- and three-point CCFs without additional constraints. This approach is developed for a 2D system of particles, but can be also used to study 3D systems provided that particles can be aligned with respect to a certain axis.

Our goal here is to determine the scattering pattern of a single particle $I_{\psi_{0}}(\mathbf{q})$  using a large number of diffraction patterns $I(\mathbf{q})$ [Eq.~(\ref{Intens1})] corresponding to different realizations of the system. Once the single-particle intensity is obtained, conventional phase retrieval algorithms \cite{Fienup, Elser} can recover the projected electron density of a single particle.
According to  Eq.~(\ref{IFSeries1}) the intensity scattered from a single particle can be uniquely determined by the set of complex coefficients $\{I_{q,\psi_{0}}^{n}\}=\{|I_{q,\psi_{0}}^{n}|,\phi_{q,\psi_{0}}^{n}=\arg(I_{q,\psi_{0}}^{n})\}$.
We will determine the Fourier components $\{I_{q,\psi_{0}}^{n}\}$ applying two- and three-point CCFs to the measured
intensities $I(q,\varphi)$ scattered from $N$ particles.

Equations~(\ref{Cq1q2Iq1Iq2A}), (\ref{PhRel1}) and (\ref{PhRel2}) constitute the core of our approach \cite{Kurta2} and allow us to
directly determine the complex Fourier components $I_{q,\psi_{0}}^{n}$.
According to our simulations, these expressions, originally derived in the dilute limit approximation [see the incoherent sum Eq.~(\ref{Iincoh1})], can be also used in the case of coherent scattering from a system of particles.
The amplitudes $\left\vert I_{q, \psi_{0}}^{n}\right\vert$ can be determined \cite{Kurta2} using Eq.~(\ref{Cq1q2Iq1Iq2A}), and the phases $\phi_{q,\psi_{0}}^{n}$ using Eqs.~(\ref{PhRel1}) and (\ref{PhRel2}), in particular when $n_{1}=n_{2}$. It might be tempting to use Eq.~(\ref{Eq:Cq1q2n_2q}) to determine the amplitudes $|I_{q,\psi_{0}}^{n}|$ by just taking the square root of  $\left\langle C_{q}^{n} \right\rangle_{M}$. However, in the case of coherent scattering from $N$ particles the spectrum $\left\langle C_{q}^{n} \right\rangle_{M}$ may contain a substantial inter-particle contribution or noise \cite{Altarelli,Kurta1}. In contrast to this, even in the case of coherent scattering from a dilute system of particles, the inter-particle contribution to the spectrum $\left\langle  C_{q_{1},q_{2}}^{n} \right\rangle_{M}$ determined at {\it different} resolution rings $q_{1} \neq q_{2}$ is negligibly small. For the same reason we used the three-point CCFs [Eq.~(\ref{Cq1q2q3})] defined  at different resolution rings $q_{1}$, $q_{2}$ and $q_{3}$. Once all amplitudes and phases of the Fourier components $I_{q, \psi_{0}}^{n}$ are determined, the diffraction pattern corresponding to a single particle can be recovered using Eq.~(\ref{IFSeries1}).

\begin{figure}[!htbp]
\centering
\includegraphics[width=0.7\textwidth]{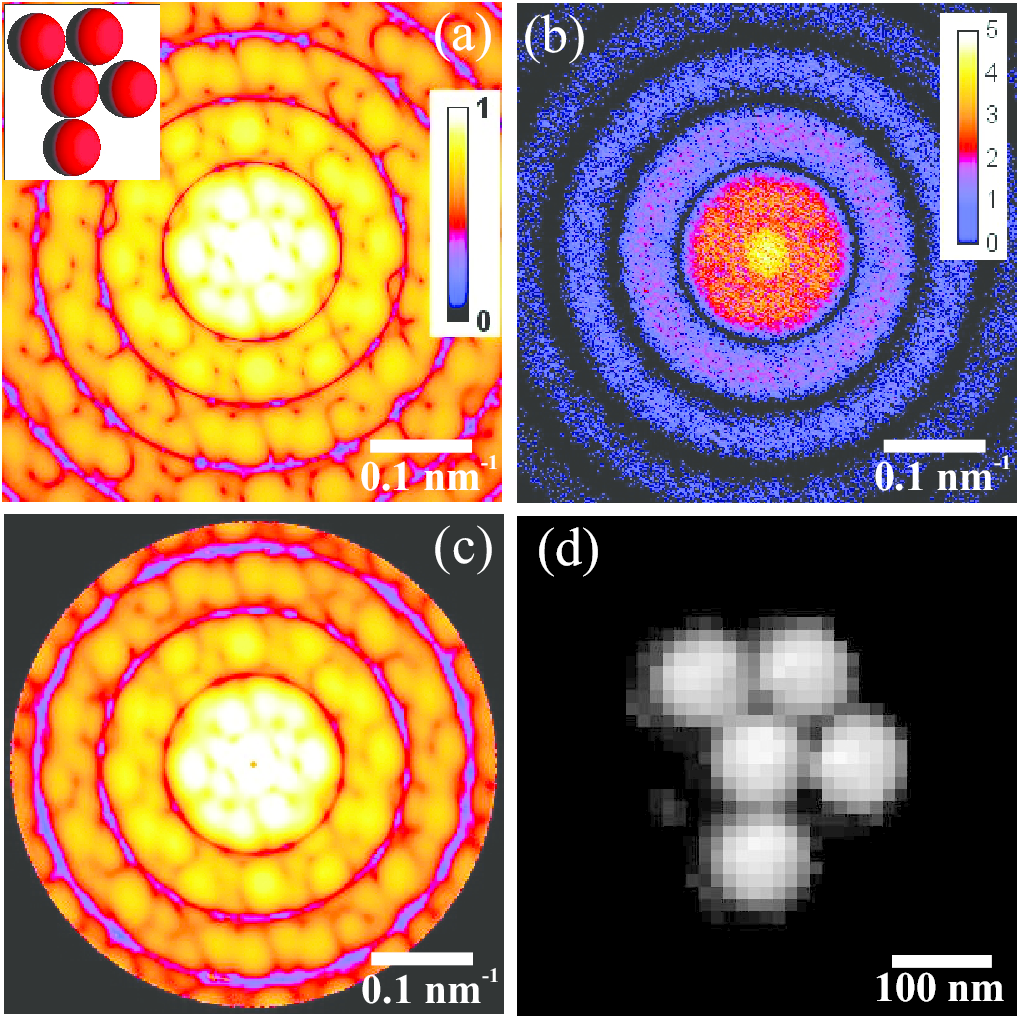}
\caption{
(a) Scattered intensity (logarithmic scale) calculated for a single asymmetric cluster (shown in the inset).
(b) Coherently scattered intensity from a disordered system consisting of $N=10$ clusters in random position and orientation. Scattered signal corresponds to the incident fluence of  $4\cdot10^{11}\;\rm{photons}/\mu\rm{m}^2$ and contains Poisson noise.
(c) Scattered intensity corresponding to an asymmetric cluster recovered from $M=10^5$ diffraction patterns of the form (b).
(d) Structure of a single cluster reconstructed by an iterative phase retrieval algorithm using the diffraction pattern shown in (c). The intensities in (a), (c) are given in arbitrary units, and in (b) in photon counts.
\label{Fig:Recovery}
}
\end{figure}

We demonstrate our approach for the case of a coherent illumination of a disordered system of particles [Eq.~(\ref{Intens1})] with an incident fluence of $4 \cdot 10^{11}\;\rm{photons}/\,\mu\rm{m}^2$ (that corresponds to
$10^{13}\;\rm{photons}$ focused on a sample area of $5\times 5 \;\mu\rm{m}^{2}$), in the presence of Poisson noise in the scattered signal. As an example, we recover a diffraction pattern and projected electron density of an asymmetric cluster [Fig.~\ref{Fig:Recovery}(a)] composed of PMMA spheres of $50\;\rm{nm}$ radius, with a size of a cluster $d=300\;\rm{nm}$. We consider coherent scattering of x-rays with wavelength $\lambda=1\;\mathring{A}$ from a system of $N=10$ clusters in random positions and orientations, distributed within the sample area [see Fig.~\ref{Fig:GeomSampl}(c)]. Diffraction patterns are simulated for a 2D detector of size $D=24\;\rm{mm}$
(with pixel size $p=80\;\mu\rm{m}$), positioned in transmission geometry at $L=3\;\rm{m}$ distance from the sample [see Fig.~\ref{Fig:GeomSampl}(a)]. This experimental geometry corresponds to scattering to a maximum resolution of $0.25\;\rm{nm}^{-1}$. For given experimental conditions the speckle size corresponding to the illuminated area is below the pixel size of the detector. At the same time the speckle size corresponding to the size of a single particle is about $12$ pixels, which provides sufficient sampling for the phase retrieval algorithm.

\begin{figure}[!ht]
\centering
\includegraphics[width=1.0\textwidth]{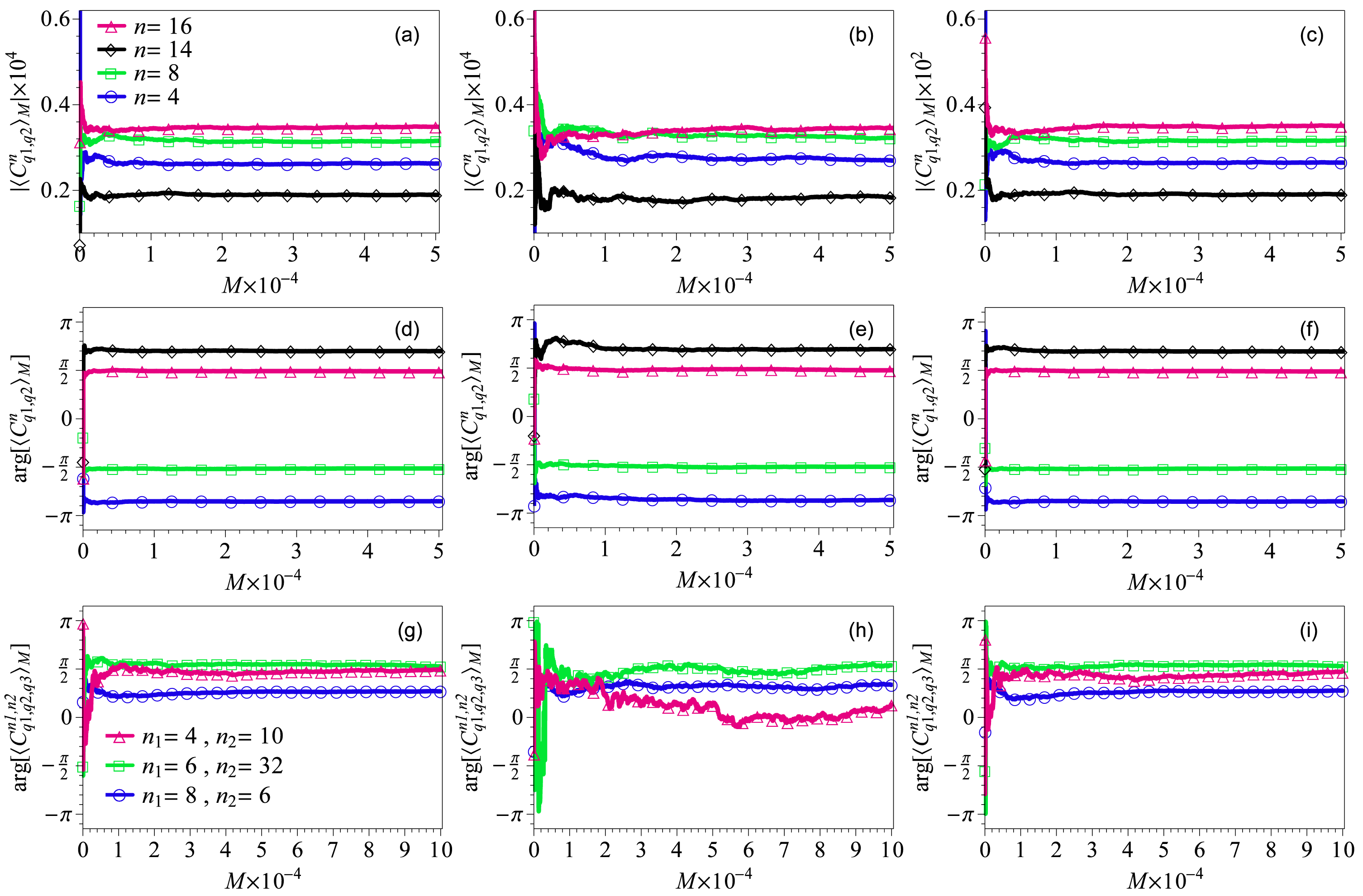}
\caption{
(a)-(f) Statistical convergence of the amplitudes $\left\vert\left\langle C_{q_{1},q_{2}}^{n}\right\rangle_{M}\right\vert$ and phases $\arg\left[\left\langle C_{q_{1},q_{2}}^{n}\right\rangle_{M}\right]$ at $q_{1}=0.25\;\rm{nm}^{-1}$, $q_{2}=0.09\;\rm{nm}^{-1}$, $n=4, 8, 14$, and $16$, calculated as a function of the number of diffraction patterns $M$ used in the averaging. (g)-(i) The phases $\arg[\left\langle C_{q_{1},q_{2},q_{3}}^{n_{1},n_{2}}\right\rangle_{M}]$ at $q_{1}=0.23\;\rm{nm}^{-1}$, $q_{2}=0.24\;\rm{nm}^{-1}$ and $q_{3}=0.25\;\rm{nm}^{-1}$, calculated for three different combinations of $n_{1}$ and $n_{2}$ as a function of $M$.
Simulations were performed (a),(d),(g) without noise, (b),(e),(h) for the incident photon fluence $4\cdot10^{10}\;\rm{photons}/\mu\rm{m}^2$, and (c),(f),(i) for the photon fluence $4\cdot10^{11}\;\rm{photons}/\mu\rm{m}^2$. In (b),(e),(h), and (c),(f),(i) Poisson noise was included.
\label{Fig:SpecConverge}
}
\end{figure}

The coherently scattered intensity simulated for a single realization of the system\footnote{All simulations of diffraction patterns were performed using the computer code MOLTRANS.} is shown in Fig.~\ref{Fig:Recovery}(b).
The Fourier components of two-point and three-point CCFs, $\left\langle C_{q_{1},q_{2}}^{n}\right\rangle_{M}$ and $\left\langle C_{q_{1},q_{2},q_{3}}^{n,n}\right\rangle_{M}$, were averaged over $M=10^5$ diffraction patterns.
It is important to perform the averaging of the two- and three-point CCFs over a sufficient number $M$ of diffraction patterns to achieve their convergence to the mean value. 
In practical applications the convergence of the CCFs can be directly controlled as a function of the number of the diffraction patterns  $M$ considered in the averaging \cite{Kurta1}. The results presented in Fig.~\ref{Fig:SpecConverge} demonstrate the evolution of the Fourier components $\left\langle C_{q_{1},q_{2}}^{n}\right\rangle_{M}$ as a function of $M$. As one can see, after averaging the CCFs over a sufficiently large number $M$ of diffraction patterns, the amplitudes and phases of $\left\langle C_{q_{1},q_{2}}^{n}\right\rangle_{M}$ asymptotically converge to their mean values. Comparing figures~\ref{Fig:SpecConverge}(a),(b) and figures~\ref{Fig:SpecConverge}(d),(e) one can see, that the CCFs converge slower in the presence of Poisson noise, and one needs to average over more patterns. This effect is more prominent for the three-point CCFs. In Fig.~\ref{Fig:SpecConverge}(g)-(i) the phases of the Fourier components $\left\langle C_{q_{1},q_{2},q_{3}}^{n_{1},n_{2}}\right\rangle_{M}$ are presented for three different combinations of $n_{1}$ and $n_{2}$.
It is well seen from these figures, that at the incident fluence of $4 \cdot 10^{11}\;\rm{photons}/\,\mu\rm{m}^2$ the phases and amplitudes of the Fourier components  are reaching their mean values at $M=10^5$.

The diffraction pattern corresponding to a single particle recovered by our approach is presented in Fig.~\ref{Fig:Recovery}(c). As one can see, it reproduces the diffraction pattern of an individual cluster shown in Fig.~\ref{Fig:Recovery}(a) very well. The structure of the cluster reconstructed from this recovered diffraction pattern was obtained by a standard phase retrieval approach \cite{Fienup} and is presented in Fig.~\ref{Fig:Recovery}(d).
The comparison of the single cluster structure obtained by our approach [Fig.~\ref{Fig:Recovery}(d)] with the initial model shown in the inset of Fig.~\ref{Fig:Recovery}(a) confirms the correctness of our reconstruction. The particle was reconstructed with a resolution of $25\;\rm{nm}$. These results clearly demonstrate the ability of our approach to recover the single-particle structure from noisy data obtained in coherent x-ray scattering experiments. A detailed discussion of different aspects that can affect the quality of reconstruction, such as particle density, fluctuations of the number of particles $N$ in the system, and noise can be found in \cite{Kurta2, Kirian}.

\subsection{Wide angle scattering from a 3D system of pentameric structures}

In the previous section we demonstrated the application of XCCA to the scattering data from a 2D disordered system of
particles which led to a successful reconstruction of the structure of an individual particle. At
the same time, for a 3D disordered system the same problem has been solved only if
additional symmetry conditions were imposed \cite{Saldin4,Starodub}. Hence, development of direct, self-consistent
approaches for the recovery of the local structure of a 3D disordered system remains a challenge.
As a step towards the solution of this problem we present the results of simulations of x-ray
scattering experiments from 3D disordered systems composed of oxygen tetrahedral pentamers [Fig.~\ref{Fig:GeomSampl}(b)].
The structure of an oxygen pentamer corresponds to the tetrahedral arrangement of water molecules
in a so-called Walrafen pentamer \cite{Walrafen}, one of the favorable models of water that reproduces the
results of x-ray and neutron scattering experiments \cite{Walrafen, Canpolat, HeadGordon, Ludwig, Wernet}.
We perform the XCCA of the diffraction patterns originating from such structures and show that the
Fourier spectra of the CCFs contain information on the local
structure of a disordered system that is not accessible in the standard small-angle x-ray scattering
(SAXS) experiment \cite{Kurta3}.

We consider here a scattering experiment on a 3D disordered system composed of
$N$ oxygen pentamers [see Fig.~\ref{Fig:GeomSampl}(a)] with the following parameters: x-ray
wavelength $\lambda=1.0\;\mathring{A}$, detector size $D = 24\;\rm{mm}$ (with a pixel size of $p = 80\;\mu\rm{m}$), sample-detector
distance $L = 25\;\rm{mm}$. The samples were composed of a different number $N$ of oxygen pentamers.
In our model we assume a uniform distribution of orientations of clusters. Three Euler angles were used to generate this distribution. The
regular pentamer [see Fig.~\ref{Fig:GeomSampl}(b)] was defined by the O-O distance of $2.82\;\mathring{A}$ between the central oxygen
atom and each atom in a vertex, and a tetrahedral angle $109.47^{\circ}$ typical for water \cite{Canpolat}.
We demonstrate results of the calculations of the Fourier spectra $\left\langle C_{q}^{n}\right\rangle_{M}$
and $\left\langle C_{q_{1},q_{2}}^{n}\right\rangle_{M}$ averaged over a large number $M$ of realizations of a system.
All results presented in this section are normalized by the number of particles $N$ considered in each particular case.

\begin{figure*}[!htbp]
\centering
\includegraphics[width=0.8\textwidth]{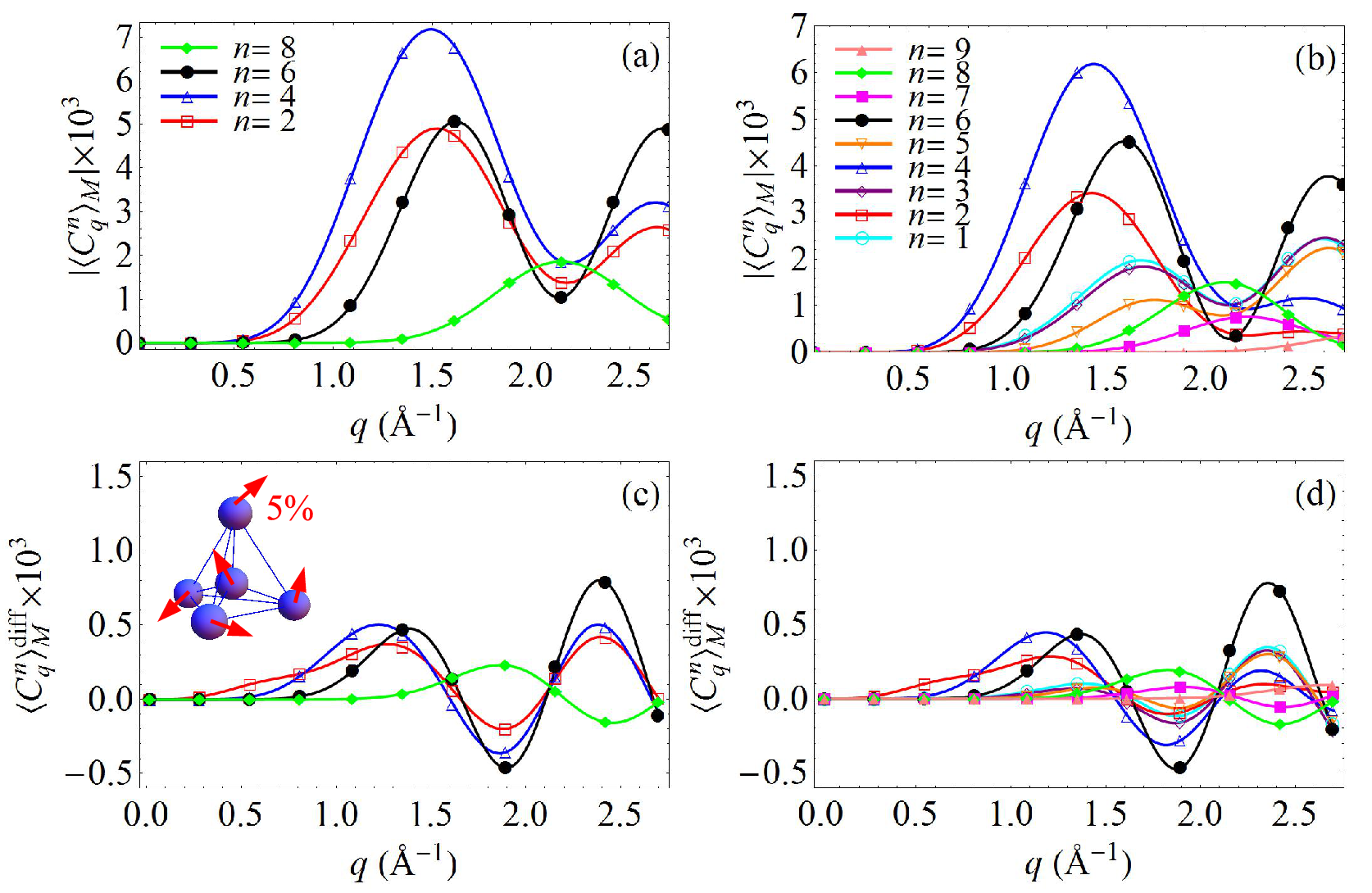}
\caption{
Fourier spectra $\left\langle C_{q}^{n}\right\rangle_{M}$ calculated as a function of $q$ for $1\leq n\leq 9$, for a flat (a),(c) and curved (b),(d) Ewald sphere.
(a),(b) Amplitudes $\left\vert\left\langle C_{q}^{n}\right\rangle_{M}\right\vert$ calculated in dilute limit approximation of scattering [Eq.~(\ref{Iincoh1})] from $N=10$ clusters, averaged over $M=10^4$ diffraction patterns.
(c),(d) Difference spectra $\left\langle C_{q}^{n}\right\rangle_{M}^{\rm{diff}}$ calculated in the dilute limit approximation for a distorted oxygen pentamer for $N=10$, and $M=10^4$ (see text for details).  All atoms in each pentamer cluster in the sample were randomly displaced by $5\%$ of the O-O distance from the positions of a regular pentamer.
\label{Fig:WaterCCF1}
}
\end{figure*}

First, we consider scattering from a dilute system of clusters, where the intensity $I(\mathbf{q})$ scattered from $N$ clusters
is approximated by a sum of intensities scattered from each cluster [see Eqs.~(\ref{Iincoh1}) and (\ref{Eq:In_k1k2_3dil})].
The Fourier spectra $\left\langle C_{q}^{n}\right\rangle_{M}$ defined on a single resolution ring $q$
and averaged over $M=10^4$ diffraction patterns for $N=10$ clusters
in the system are presented in Figs.~\ref{Fig:WaterCCF1}(a)-(b). The spectrum $C_{q}^{n}$ for each realization of the system
was directly determined using Eqs.~(\ref{Eq:Cq1q2n_2q}) and (\ref{Eq:In_k1k2_3dil}).
Calculations were performed for a regular pentamer for the
case of a flat (a) and curved (b) Ewald sphere.
Calculations for the case of a flat Ewald sphere \footnote{Such conditions can be achieved, for example, using high photon energy.}
were performed by setting the $z$-component of the momentum transfer vector to $q^{z}=0$ in Eq.~(\ref{Eq:In_k1k2_3dil}).
In the case of a flat Ewald sphere only Fourier components with even $n$-values have nonzero values.
The amplitudes $\left\vert\left\langle C_{q}^{n}\right\rangle_{M}\right\vert$ calculated for a single cluster
$N=1$ in the system (not presented here), and for a system containing $N=10$ clusters [Figs.~\ref{Fig:WaterCCF1}(a) and (b)], converge to the
same functional dependence of $q$. Therefore, in dilute limit approximation the results do not depend on the number $N$ of particles
in the system, but are just scaled by $N$. As one can see from Figs.~\ref{Fig:WaterCCF1}(a)-(b),
in the considered $q$-range only the Fourier components with $n\leq 9$ contribute to the spectrum, and in the case of the curved Ewald sphere the Fourier componets
with odd $n$ have magnitudes comparable with the ones with even $n$. Therefore, experimental conditions that correspond to curved Ewald sphere can provide additional
structural information, as compared to the case with a flat Ewald sphere.

The effect of a distortion of the regular pentamer on the Fourier spectra of the two-point CCF is demonstrated
in Figs.~\ref{Fig:WaterCCF1}(c),(d). These results were also obtained in the dilute limit using Eq.~(\ref{Eq:In_k1k2_3dil}).
The difference spectra $\left\langle C_{q}^{n}\right\rangle_{M}^{\rm{diff}}=\left\vert\left\langle C_{q}^{n}\right\rangle_{M}^{\rm{dist}}\right\vert-\left\vert\left\langle C_{q}^{n}\right\rangle_{M}^{\rm{reg}}\right\vert$
were calculated for the case when all atoms in each cluster were randomly displaced by $5\%$ of the O-O distance. Here,
$\left\langle C_{q}^{n}\right\rangle_{M}^{\rm{reg}}$ corresponds to the spectra calculated for the regular pentamer [Figs.~\ref{Fig:WaterCCF1} (a), (b)],
and $\left\langle C_{q}^{n}\right\rangle_{M}^{\rm{dist}}$ to the distorted one (not presented). 
These results show that CCFs preserve their functional dependence on momentum transfer vector $q$ with small distortions of the cluster, and, importantly, the ratio between the values of different Fourier components is not significantly changed.

\begin{figure*}[!htbp]
\centering
\includegraphics[width=0.8\textwidth]{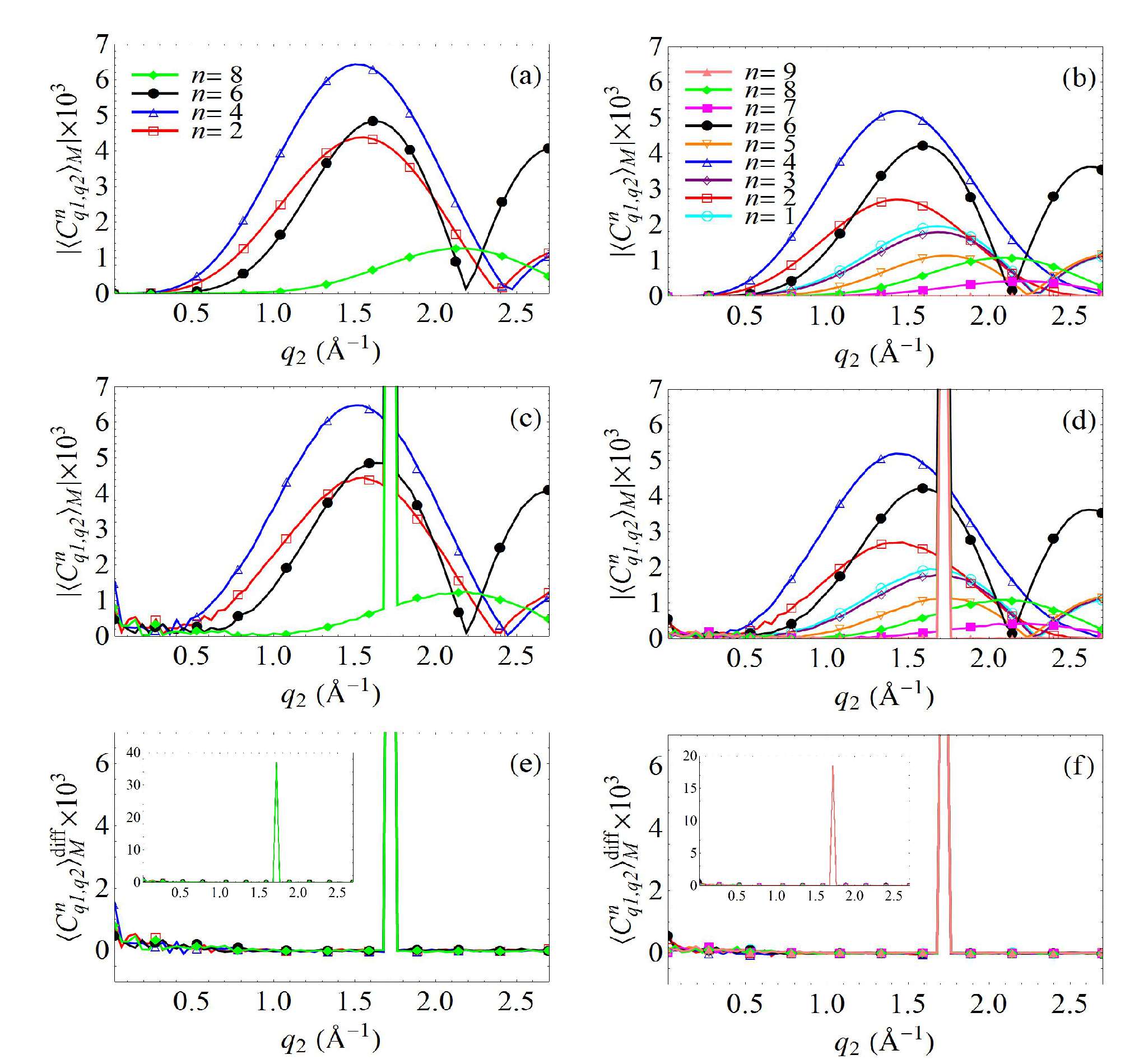}
\caption{Fourier spectra $\left\langle C_{q_{1},q_{2}}^{n} \right\rangle_{M}$ calculated as a function of $q_{2}$ ($q_{1}=1.71\;\mathring{A}^{-1}$) for $1\leq n\leq 9$, for a flat (a),(c),(e)  and curved (b),(d),(f) Ewald sphere. (a),(b) Amplitudes $\left\vert\left\langle C_{q_{1},q_{2}}^{n} \right\rangle_{M}\right\vert$ calculated in dilute limit approximation of scattering [see Eq.~(\ref{Iincoh1})] from $N=10$ clusters, averaged over $M=10^4$ diffraction patterns.
(c),(d)  Amplitudes $\left\vert\left\langle C_{q_{1},q_{2}}^{n} \right\rangle_{M}\right\vert$ calculated for the case of coherent scattering [Eq.~(\ref{Intens1})]
from a cubic sample of $800 \;\rm{nm}$ in size consisting of $N=400$ clusters, averaged over $M=10^6$ diffraction patterns.
(e),(f) Difference spectra $\left\langle C_{q_{1},q_{2}}^{n} \right\rangle_{M}^{\rm{diff}}$
show the difference (e) between the spectra in (c) and (a), and (f) between (d) and (b). The insets in (e) and (f) show full size of the peaks attributed to the inter-particle contribution.
\label{Fig:WaterCCF2}
}
\end{figure*}

It is interesting to consider similar results of simulations for the Fourier spectra $\left\langle C_{q_{1},q_{2}}^{n} \right\rangle_{M}$ defined at different resolution rings $q_{1} \neq q_{2}$.
In Figs.~\ref{Fig:WaterCCF2}(a)-(d) the amplitudes $\left\vert\left\langle C_{q_{1},q_{2}}^{n} \right\rangle_{M}\right\vert$
calculated as a function of $q_{2}$ ($q_{1}=1.71\;\mathring{A}^{-1}$) are presented for a flat (a),(c) and curved (b),(d) Ewald sphere.
The amplitudes $\left\vert\left\langle C_{q_{1},q_{2}}^{n} \right\rangle_{M}\right\vert$ shown in Figs.~\ref{Fig:WaterCCF2}(a),(b) were calculated
in dilute limit from $N=10$ clusters, using Eq.~(\ref{Eq:In_k1k2_3dil}) in (\ref{Eq:Cq1q2n_2}), and were averaged [Eq.~(\ref{AverCorrSpec1})] over $M=10^4$ realizations of the system.
The spectra shown in Figs.~\ref{Fig:WaterCCF2}(c),(d) were calculated for the case of coherent scattering
from a cubic sample of $800 \;\rm{nm}$ in size consisting of $N=400$ clusters.
In this case the coherently scattered intensities [Eq.~(\ref{Intens1})] were calculated for each realization of the system,
and then the two-point CCFs and their Fourier components were determined using Eqs.~(\ref{Eq:Cq1q2_4}) and (\ref{Eq:Cq1q2n_1}) and averaged [Eq.~(\ref{AverCorrSpec1})] over $M=10^6$ diffraction patterns.
A much larger number of diffraction patterns was considered in the latter case to average out the inter-particle contribution in the specra, which is absent in the former case.
The difference between the spectra in Figs.~\ref{Fig:WaterCCF2}(c) and ~\ref{Fig:WaterCCF2}(a), and in Figs.~\ref{Fig:WaterCCF2}(d) and ~\ref{Fig:WaterCCF2}(b), is presented in Figs.~\ref{Fig:WaterCCF2}(e) and \ref{Fig:WaterCCF2}(f), respectively.
It is easy to see from Fig.~\ref{Fig:WaterCCF2}, that the inter-particle contribution to the Fourier components $\left\langle C_{q_{1},q_{2}}^{n} \right\rangle_{M}$
in the case of coherent scattering from $N=400$ clusters is negligible in the entire $q$-range, except of a small region corresponding to $q_{1}=q_{2}=1.71\;\mathring{A}^{-1}$.
This is reflected in a sharp peak in the difference spectra presented in Figs.~\ref{Fig:WaterCCF2}(e),(f). The width of this peak is related to the speckle size given by the sample size.
As one can see from the insets in Figs.~\ref{Fig:WaterCCF2}(e) and \ref{Fig:WaterCCF2}(f), the magnitude of the inter-particle contribution at $q_{1}=q_{2}$ is much higher than the structural contribution.
This clearly shows, that in the case of coherent scattering from a dilute system of particles,
the inter-particle contribution to the Fourier components $\left\langle C_{q_{1},q_{2}}^{n} \right\rangle_{M}$
defined at $q_{1} \neq q_{2}$ is negligibly small in contrast to the case $q_{1}=q_{2}$, where this contribution can be very large \cite{Altarelli, Kurta1}.

Results presented here demonstrate the ability of XCCA to obtain information about the oxygen
clusters with tetrahedral arrangement of oxygen atoms typical for water. The spectra $\left\langle C_{q_{1},q_{2}}^{n} \right\rangle_{M}$ averaged for a large number of realizations of a system
converge to a specific functional dependence of $q$, which one could hope to correlate with the local structure of the 3D system composed of many particles.
Our results show, that small distortions of the oxygen pentamer do not lead to significant changes of the
CCFs. This may help to identify an average local structure of water using experimentally determined
CCFs. Our results demonstrate that the spectra $\left\langle C_{q_{1},q_{2}}^{n} \right\rangle_{M}$ defined at $q_{1} \neq q_{2}$
are more suitable for investigation of the structure of disordered system in the case of coherent scattering, than the spectra determined at $q_{1} = q_{2}$.
The Fourier components $\left\langle C_{q_{1},q_{2}}^{n} \right\rangle_{M}$  can provide the structural information in the case when
it is suppressed by a large inter-particle background in $\left\langle C_{q}^{n}\right\rangle_{M}$. Our results show, that the Fourier components
$\left\langle C_{q_{1},q_{2}}^{n} \right\rangle_{M}$ can reveal structural information about local structure of a system with a larger number of particles,
as compared to $\left\langle C_{q}^{n}\right\rangle_{M}$.


\section{Conclusions}

In summary, we showed here how XCCA can be applied to study the local structure of disordered materials. Important steps in this analysis are Fourier decomposition of CCFs and averaging over many realizations of the system. As it was shown in our work, statistical fluctuations of the Fourier components of CCFs, due to this averaging cancel out and information on the local structure of the system can be deduced. The most transparent results can be obtained in the analysis of the dilute 2D systems when contribution from the inter-particle correlations can be neglected. We demonstrated that in this specific case all Fourier coefficients of the angular decomposition of the coherently scattered intensity of a single unit can be completely recovered by the analysis of two- and three-point correlation functions. Importantly, not only amplitudes but also phases of this decomposition can be determined directly from the experimental data. Applying standard phase retrieval techniques to the recovered intensity gives the structure of a single unit in a disordered system.

The XCCA of dense systems shows more complicated features in the Fourier components of CCFs due to the presence of a substantial inter-particle contribution. It can be significantly reduced in some situations when CCFs are analyzed at different resolution rings. Our analysis showed that inter-particle correlations may strongly influence the values of Fourier components of CCFs defined at the same resolution ring.  An extraction of a single-particle structure in dense systems by XCCA can be more complicated, however, it may provide information on the medium range order.

The generalization of this approach to 3D systems is still a challenge. Here scattering to high angles can be especially interesting due to the presence of a significant Ewald sphere curvature. Such scattering conditions allow to determine odd Fourier components of CCFs, which are inaccessible in a standard small-angle x-ray scattering experiment, and in this way explore hidden symmetries in the 3D system. This approach can be especially attractive for the study of atomic systems.

We foresee that the general approaches described here for the analysis of the structure of disordered systems based on the angular cross-correlation techniques will find their wide applications in future. They could become especially attractive at a newly emerging free-electron lasers and diffraction limited ultimate storage rings \cite{Bei}.


\section{Acknowledgments}

We are thankful to E.~Weckert for a permanent interest and support of this project and to U. Lorenz and A. Singer for a careful reading of the manuscript. Part of this work was supported by BMBF Proposal 05K10CHG `'Coherent Diffraction Imaging and Scattering of Ultrashort Coherent Pulses with Matter`' in the framework of the German-Russian collaboration `'Development and Use of Accelerator-Based Photon Sources`' and
the Virtual Institute VH-VI-403 of the Helmholtz association.


\bibliography{XCCA_Review}

\newpage


\end{document}